\documentclass[10pt]{article}
\usepackage{epsfig,float}
\usepackage{amsmath}

\begin{document}
\title{Stochastic resonance as a collective property of ion channel assemblies}
\author{G. Schmid, I. Goychuk, and P. H\"anggi\\ {\small University of
    Augsburg, Department of Physics, D-86135 Augsburg, Germany}}

\maketitle

\begin{abstract}
By use of a stochastic generalization of the
Hodgkin-Huxley model we  investigate both the phenomena of {\it
stochastic resonance} (SR) and  {\it coherence resonance} (CR) in  variable
size patches of an excitable cell membrane. Our focus is on the challenge how
internal noise stemming from
individual ion channels does affect collective properties of the whole
ensemble. We investigate both  an unperturbed 
 situation with no applied stimuli and one in which the membrane
 is stimulated externally by a
 periodic signal and additional external noise. For the nondriven case, we
demonstrate the existence of an  optimal size of the membrane patch for which 
the
{\it internal noise}  causes a most regular  spike activity. This phenomenon
shall be termed {\it intrinsic} CR. In presence of an applied periodic
stimulus  we demonstrate that the signal-to-noise ratio (SNR)  exhibits 
SR {\it vs.}\/ decreasing patch size, or  {\it vs.} 
increasing {\it internal} noise strength, respectively.
Moreover, we demonstrate that conventional SR {\it vs.\/}  the
{\it external} noise intensity occurs
only for sufficiently large membrane patches, when the intensity of internal
noise is below its optimal level.  Thus,  biological SR  seemingly is rooted in 
the {\it collective}
properties of large ion channel ensembles rather than in the
individual stochastic dynamics of single ion channels.  
\end{abstract}

\noindent
PACS number(s): 87.10.+e, 87.16.-b, 05.40.-a

During the last decade, the effect of stochastic resonance (SR) -- 
a cooperative
phenomenon wherein the addition of external noise improves the 
detection and
transduction of signals in nonlinear systems (for a comprehensive survey and
 prominent references, see Ref. \cite{Gammaitoni98}) -- 
has been studied  experimentally and theoretically in various biological systems
\cite{Douglass93,Levin96,Collins96,Simonotto97,Hidaka00}. 
For example, SR has been experimentally demonstrated within the 
mechanoreceptive system in crayfish
\cite{Douglass93}, in the cricket cercal sensory system \cite{Levin96}, for
 human tactile sensation \cite{Collins96}, visual perception
\cite{Simonotto97}, and response behavior of the arterial baroreflex system of humans 
\cite{Hidaka00}. The importance of this SR-phenomenon for sensory biology
is by now   well established; yet, it is presently not known to which minimal level
of the biological organization the stochastic resonance effect can ultimately
be traced down.
Presumably, SR has its origin in the stochastic properties of
the ion channel clusters located in a receptor cell membrane. Indeed, for an
artificial model system
Bezrukov and Vodyanoy have demonstrated experimentally that a large parallel 
ensemble of the alamethicin ion channels does exhibit stochastic resonance
\cite{Bezrukov95}. This in turn provokes the question whether a
{\it single}\/ ion channel is able to exhibit SR, or whether 
stochastic resonance
is the result of a {\it collective} response from a finite assembly
of channels. 
Stochastic resonance in single, biological potassium ion channels 
has also been investigated both theoretically \cite{GH00} and 
experimentally \cite{petracchi}. This very experiment did 
not convincingly exhibit 
 SR in single voltage-sensitive 
ion channels.  Nevertheless, the SR phenomenon  can occur 
in a single ion channel if only the parameters are within a regime where the
channel is predominantly dwelled in the closed state, as demonstrated
  within a theoretical modeling for a potassium Shaker channel 
  \cite{GH00}.
 This prominent result, {\it i.e.} the manifestation of SR on the {\it single}-molecular level,
 is not only of academic 
interest, but is relevant also 
for potential nanotechnological applications,
such as the  design of single-molecular biosensors. 
 The origin and  biological relevance  of SR in single ion
channels, however, remains still open. Indeed,  biological SR can be a
manifestation of  {\it collective} properties of  large assemblies of ion
channels of different sorts. To display the phenomenon of 
excitability 
these assemblies must contain a collection of  ion channels of at least two
different kinds -- such as, {\it e.g.},  potassium- and sodium-channels. The corresponding
mean-field type model has been put forward by Hodgkin and Huxley in
1952 \cite{Hodgkin52} by neglecting the mesoscopic fluctuations
which originate from the stochastic opening and closing of channels. 
SR due to {\it external}
noise in this primary model and related
 models of excitable dynamics has
extensively been addressed \cite{SRneuron}. These 
models further display
 another interesting effect in the presence of noise, namely so termed
coherence resonance (CR) \cite{Pikovsky97}: even in absence of an external
periodic signal the stochastic dynamics exhibits a surprisingly
 more regular behavior solely due to an
optimally applied external noise intensity. A challenge though still
remains: Does {\it internal} noise play a constructive role for SR and CR? 
Internal noise is produced by fluctuations
of individual channels within the assembly, and diminishes with increasing number. 
For a large, macroscopic number of channels this noise becomes negligible. 
Under the realistic biological conditions, however, it may play a crucial role.

Our starting point is due  to the model of Hodgkin and Huxley
\cite{Hodgkin52}, {\it i.e.}
the ion current across the biological membrane is carried mainly by
the motion of sodium, {Na$^{+}$}, and potassium, {K$^{+}$}, 
ions through 
selective and
voltage-gated ion channels embedded across the membrane. Besides,
there is also a leakage current present which is  induced by
chloride and remaining other ions. The ion channels are formed
by special membrane proteins  which undergo spontaneous, but
voltage-sensitive
conformational transitions between open and closed states \cite{Hille92}.
Moreover, the conductance of the
membrane is directly proportional to the number of the {\it open} ion channels.
This number depends on the potential difference across the membrane,
$V$. The different concentrations of the ions inside and outside the cell are
encoded by corresponding  reversal potentials 
$E_{\rm Na}=50{\rm mV}$, $E_{K}=-77{\rm mV}$ and $E_{{\rm L}}=-54.4{\rm mV}$, respectively, 
 which give the
voltage values at which the direction of partial ion currents is reversed 
\cite{Hille92}. Taking into account that
 the membrane possesses a
capacitance $C=1{\rm \mu F/cm^{2}}$, 
Kirchhoff's first law reads in presence
of an  {\it external} current $I_{{\rm ext}}(t)$  stimulus:

\begin{align}
\label{voltage-equation}
  C \frac{d}{dt} V +G_{{\rm K}}(n)\ (V-E_{{\rm K}})+G_{{\rm Na}}(m,h)\ ( V - E_{{\rm Na}}) +G_{{\rm L}}
\ (V - E_{L}) = I_{{\rm ext}}(t)\, . 
\end{align}
Here, 
$G_{{\rm Na}}(m,h)$, $G_{{\rm K}}(n)$ and $G_{{\rm L}}$ denote the conductances
of  sodium, potassium,  and the remaining other ion channels, respectively.
The leakage conductance is assumed to be constant, $G_{{\rm L}}=0.3{\rm mS/cm^{2}}$;
in contrast, those of sodium and potassium depend on the probability to find the
ion channels in their open conformation.
To explain the experimental data, Hodgkin and Huxley did assume that the
conductance of a potassium channel is gated by  four independent 
and identical gates.
Thus, if $n$ is the probability of one gate to be open, the probability
of the K$^{+}$ 
channel to stay open is $P_{{\rm K}}=n^4$. 
Moreover, the gating dynamics of sodium channel is assumed to be governed
by  three independent, identical gates with opening probability $m$ 
and an additional one, being different,
possessing the opening probability $h$. 
Accordingly, the
opening probability of Na$^{+}$ channel (or the fraction of open
channels) reads $P_{{\rm Na}}=m^3h$. The conductances for potassium and sodium thus read
 
\begin{align}
\label{conductances-hodgkinhuxley}      
G_{{\rm K}}(n)&=g_{{\rm K}}^{{\rm max}}\ n^{4} , \quad
G_{{\rm Na}}(m,h)=g_{{\rm Na}}^{{\rm max}}\ m^{3} h\, ,
\end{align}
where $g_{{\rm Na}}^{{\rm max}}=120{\rm mS/cm^{2}}$ and $g_{{\rm K}}^{{\rm max}}=36{\rm mS/cm^{2}}$ are the
maximal conductances. 
Furthermore, the gating variables (probabilities) 
$m$, $h$ and $n$ obey the two-state, "opening-closing"
dynamics,
\begin{align}
\label{m-element}
\dot{m} &= \alpha_{m}(V)\ (1-m)-\beta_{m}(V)\ m\, ,  \nonumber \\
\dot{h} &= \alpha_{h}(V)\ (1-h)-\beta_{h}(V)\ h\, , \\
\dot{n} &= \alpha_{n}(V)\ (1-n)-\beta_{n}(V)\ n\, , \nonumber
\end{align}
with the experimentally determined voltage-dependent transition rates,
 reading \cite{Hodgkin52,nossal}
\begin{align}\label{cc}
\alpha_{m}(V) = \frac{ 0.1 ( V + 40 )}{1-\exp [ - ( V + 40 ) / 10] },\quad 
& \beta_{m}(V) = 4 \ \exp [ - ( V + 65 ) / 18 ]\, , \nonumber \\
\alpha_{h}( V ) =  0.07 \ \exp [ - ( V + 65 ) / 20 ],\quad 
& \beta_{h}( V ) = \{ 1 + \exp [ - ( V + 35 ) / 10 ] \}^{-1}\, , \\
\alpha_{n}( V ) = \frac{ 0.01 \ ( V + 55 )}{ 1 - \exp [ -( V + 55 )/10 ]},\quad 
& \beta_{n}( V ) = 0.125 \ \exp [ - ( V + 65 ) / 80 ]\, . \nonumber
\end{align}
These nonlinear Hodgkin-Huxley equations (\ref{voltage-equation})-(\ref{cc})
present a  cornerstone model in  neurophysiology. Within
the same line of reasoning this model can be generalized to 
a mixture of different ion channels with various gating 
properties  \cite{nossal,lowen99}.
An essential drawback of this model, however, is that it operates
with the {\it average} number of open channels, thereby disregarding  
corresponding
number fluctuations (or, the so-called {\it channel noise} \cite{white00}). 
Thus, it can strictly be valid only within the 
limit of very large system
size. We emphasize, however, that the size of an excitable
 membrane patch
within a  receptor cell is realistically finite.
As a consequence, the role of internal fluctuations cannot be {\it a priori}\/
neglected; as a matter of fact, as shown below, 
they do play a key role for SR and CR.

The role of channel noise for the neuron firing has been 
studied 
by Lecar and Nossal as early as in 1971 \cite{lecar}.
The corresponding stochastic generalizations of Hodgkin-Huxley model
(within a kinetic model which corresponds to the above given
description)
has been put forward by DeFelice {\it et al.} \cite{DeFelice}
and others; see \cite{white00} for a review and further references
therein. The main conclusion of these previous
studies is that the channel noise can be functionally important
for neuron dynamics. It particular, it has been demonstrated that 
channel noise alone can give rise to a spiking activity even 
in the absence of any stimulus \cite{white00,DeFelice}.

The stochastic kinetic scheme \cite{white00,DeFelice}, however, 
necessitates extensive numerical 
simulations \cite{Jung}.
To aim at a less cumbersome numerical scheme we use a short-cut
procedure that starts from  Eq. (\ref{m-element})
in order to  derive a corresponding set of Langevin equations for a stochastic
generalization of the Hodgkin-Huxley equations 
of the type put forward by Fox and Lu \cite{Fox94}. 
Following their reasoning  we substitute
the equations (\ref{m-element}) with the corresponding 
Langevin generalization: 

\begin{align}
\label{n-element-fox-lu}
    \dot{m} &= \alpha_{m}(V)\ (1-m)-\beta_{m}(V)\ m + \xi_{m}(t)\, , \nonumber \\
    \dot{h} &= \alpha_{h}(V)\ (1-h)-\beta_{h}(V)\ h + \xi_{h}(t)\, , \\
    \dot{n} &= \alpha_{n}(V)\ (1-n)-\beta_{n}(V)\ n + \xi_{n}(t)\, , \nonumber       
\end{align}
with  independent Gaussian white noise sources of vanishing mean.
 The noise autocorrelation functions depend 
on the stochastic voltage and the corresponding total 
number of ion channels as follows:

\begin{align}
\label {fox-lu-noise}
    \langle \xi_{m}(t) \xi_{m}(t') \rangle &= \frac{2}{N_{{\rm Na}}}\ \frac{ \alpha_{m}
      \beta_{m}}{(\alpha_{m} +\beta_{m})}\ \delta(t -t')\, , \nonumber \\
    \langle \xi_{h}(t) \xi_{h}(t') \rangle &=  \frac{2}{N_{{\rm Na}}}\ \frac{ \alpha_{h}
      \beta_{h}}{(\alpha_{h} +\beta_{h})}\ \delta(t -t')\, , \\
    \langle \xi_{n}(t) \xi_{n}(t') \rangle &=  \frac{2}{N_{{\rm K}}}\ \frac{ \alpha_{n}
      \beta_{n}} {(\alpha_{n} +\beta_{n})}\ \delta(t -t') \nonumber \, . 
\end{align}
In order to confine the conductances between the physically allowed values between
  $0$ (all
channels closed) and $g^{{\rm max}}$ (all channels open) we have 
implemented numerically
the constraint of reflecting boundaries so 
that $m(t)$, $h(t)$ and $n(t)$ are always located 
between zero and one \cite{Fox94}.

Moreover, the numbers  $N_{{\rm Na}}$ and $N_{{\rm K}}$
depend on the actual  area $S$ of the membrane patch. With the assumption of homogeneous
 ion channels
densities, $\rho_{{\rm Na}}=60{\rm \mu m^{2}}$ and $\rho_{{\rm K}}=18{\rm \mu m^{2}}$,  
the following scaling behavior follows:
\begin{align}
        N_{{\rm Na}}= \rho_{{\rm Na}} S, \quad N_{{\rm K}}= \rho_{{\rm K}}
        S\, .
\end{align}
 Upon  decreasing the system 
size $S$, the fluctuations and, hence, the internal noise 
increases.   

Before integrating the system of stochastic equations
(\ref{voltage-equation}), (\ref{n-element-fox-lu}), (\ref{fox-lu-noise})
numerically, the external stimulus $I_{{\rm ext}}(t)$ must be specified.
We  take a periodic stimulus of the form
\begin{align}
\label{external-stimulus}
I_{{\rm ext}}(t) = A\ \sin( \Omega t ) + \eta(t)\, ,
\end{align}
where the sinusoidal signal with amplitude $A$ and frequency $\Omega$
is contaminated by the 
Gaussian white noise $\eta(t)$\/ with the autocorrelation function
\begin{align}
\langle  \eta(t) \eta(t') \rangle =  2 D_{{\rm ext}} \ \delta(t -t')\, ,
\end{align}
and  noise strength $D_{{\rm ext}}$. The numerical integration is carried out by 
the standard Euler algorithm with the step size $\Delta t\approx
2\cdot 10^{-3}{\rm ms}$. The "Numerical Recipes" routine {\texttt
  ran2} is used for the generation of independent random numbers
\cite{Numerical-Recipes} with
the Box-Muller algorithm providing the Gaussian distributed random numbers.    
The total integration time is chosen to be a multiple of the
driving period $T_{\Omega}=2\pi / \Omega$, as to
ensure that the spectral line of the driving signal is centered on a
computed value of the power spectral densities. From the stochastic 
voltage signal $V(t)$
we extract 
a point process of spike occurrences $\{t_{i}\}$:
\begin{align}
u(t) := \sum_{i=1}^{N} \delta( t - t_{i} )\, ,
\end{align}
where $N$ is the total number of spikes occurring during the elapsed time interval. 
The occurrence of a spike
in the voltage signal $V(t)$ is obtained by upward-crossing a certain
detection threshold value $V_{0}$. It turns out that the threshold can be varied over a
wide range with no effect on the resulting spike train dynamics. However, to account
for the typical spike duration a time interval 
of $2{\rm ms}$ has been used.
The power spectral density of the spike train (PSD$_{u}$), the
interspike interval histogram (ISIH) and the coefficient of variation
(CV) have been analyzed. The coefficient of variation CV, which
presents a measure of the spike coherence, reads:
\begin{align}
  \label{coefficient-of-variation}
  {{\rm CV}} := \frac{\sqrt{\langle T^{2} \rangle -  \langle T
      \rangle^{2}}}{\langle T \rangle}\, ,
\end{align}
where $\langle T \rangle := \lim_{N \to \infty} \frac{1}{N}\sum
(t_{i+1}-t_{i})$ and
$\langle T^{2} \rangle := 
\lim_{N \to \infty} \frac{1}{N}\sum (t_{i+1}-t_{i})^{2}$  
are the mean and mean-squared interspike intervals, respectively.  From the PSD$_{u}$ 
we obtain the height of the spectral line of the
periodic stimulus as the difference between the peak value and its
background offset.  
The signal-to-noise ratio (SNR) is then given by the ratio of 
signal peak height to the background height 
(in the units of spectral resolution of signals).  

\begin{figure}[t]
  \centerline{\epsfxsize=0.9\textwidth \epsffile{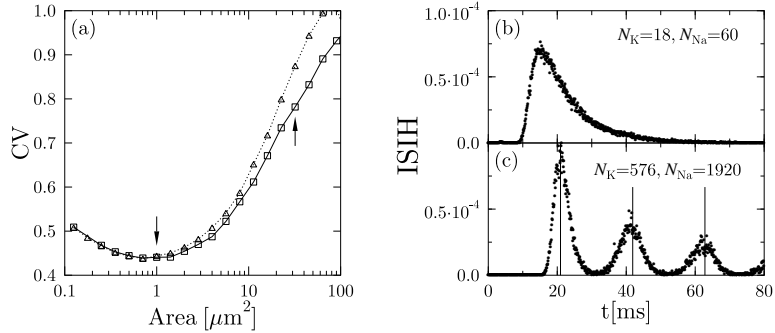}}
  \caption{   
    The CV defined in equation
    (\ref{coefficient-of-variation}) plotted versus area $S$ (a) for
    $A=1.0{\rm \mu A/cm^{2}}$,
    $\Omega=0.3{\rm ms^{-1}}$, in the absence of external noise
    $D_{{\rm ext}}=0$ (solid line) and with no external stimulus 
    applied (dotted line). 
    The ISIH are depicted in the presence of signal for
    $S=1{\rm \mu m^{2}}$ (b),  and $S=32{\rm \mu m^{2}}$ (c). The vertical lines
    denote the driving period and the first two multiples. }
  \label{coherence-no-external-noise}
\end{figure}

\begin{figure}[t]
  \centerline{\epsfxsize=0.9\textwidth \epsffile{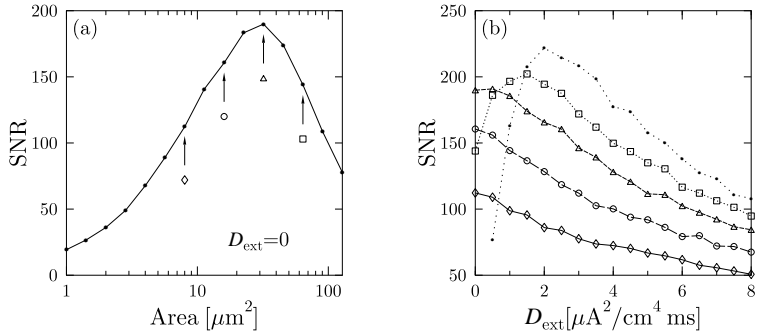}}
  \caption{  
    The signal-to-noise ratio (SNR) for an external
    sinusoidal stimulus with amplitude $A=1.0{\rm \mu A/cm^{2}}$ and frequency
    $\Omega=0.3{\rm ms^{-1}}$ for different observation areas: (a) 
    No external noise is applied; (b) SNR versus the external noise
    for the system sizes indicated by the arrows in fig.~\ref{snr-spike-psd}(a). 
    $S=8{\rm \mu m^{2}}$ ($N_{{\rm K}}=144$, $N_{{\rm Na}}=480$): solid line through
    the diamonds, 
    $S=16{\rm \mu m^{2}}$ ($N_{{\rm K}}=288$, $N_{{\rm Na}}=960$): long dashed line
    connecting the circles, 
    $S=32{\rm \mu m^{2}}$ ($N_{{\rm K}}=576$, $N_{{\rm Na}}=1920$): short dashed line
    through the triangles, 
    $S=64{\rm \mu m^{2}}$ ($N_{{\rm K}}=1152$, $N_{{\rm Na}}=3840$): dotted line
    connecting the squares. The situation with no internal noise
    ({\it i.e.}, formally $S\to \infty$) is depicted by the dotted line.}
  \label{snr-spike-psd}
\end{figure}

We have analyzed the spike coherence in the autonomous, nondriven 
regime ({\it i.e.}, we use $I_{{\rm ext}}=0$) as a function of 
the decreasing cluster size. Our
simulation reveals, cf. fig.~\ref{coherence-no-external-noise}(a), the novel phenomenon of 
{\it intrinsic coherence resonance} , 
where the order in the spike sequence increases; {\it i.e.} the CV is decreasing, 
{\it solely due to the presence of internal noise}. 
The fully disordered sequence 
(which corresponds to a Poissonian spike train) would assume the value ${{\rm CV}}=1$.
We note, however, 
 that near $S=1{\rm \mu m^2}$ (optimal dose of internal noise), 
${{\rm CV}}\approx 0.44$, {\it i.e.} the spike train becomes distinctly more ordered!
For $S<1{\rm \mu m^2}$, the internal noise increases further 
beyond the optimal value
and destroys the order in spiking again. It is worth noting  that for 
$S<1{\rm \mu m^2}$ the model reaches limiting validity; in that regime
 the  kinetic scheme \cite{white00,DeFelice,Jung} should be used instead.
Such a corresponding study, however, has been put forward independently
 by Jung and Shuai \cite{Jung}; their results
are in qualitative agreement with our findings. Next we switch on an external
 sinusoidal driving: Interestingly enough 
 the interspike intervals distribution is 
not affected for small patch sizes, cf. fig.~\ref{coherence-no-external-noise}(b) for $S=1{\rm \mu m^2}$. 
In this case, 
the spike-activity possesses an internal rhythm which 
dominates over the external disturbances.
For larger patch sizes the internal noise decreases and 
the periodic drive induces a reduction 
of the CV as compared to the undriven case, note the 
right arrow in fig.~\ref{coherence-no-external-noise}(a).
In this latter regime the external driving rules 
the spiking activity as depicted with the characteristic peaks in the ISIH in
fig.~\ref{coherence-no-external-noise}(c) at multiple driving periods.

Next, the focus is  on the SNR in absence of external noise,
see fig.~\ref{snr-spike-psd}(a). 
Here we discover the novel effect of genuine {\it intrinsic stochastic
resonance}, where the response of the system to the external stimulus
is optimized {\it solely} due to internal, omnipresent noise. For the given 
parameters it occurs at $S\approx 32{\rm \mu m^2}$. For $S< 32{\rm \mu m^2}$
 growing internal noise monotonically deteriorates the system response.
Under such circumstances, one would predict that the addition of an external 
noise (which corresponds to the conventional  situation
in  biological SR studies) {\it cannot}  improve SNR further,
{\it i.e.} conventional SR will not be exhibited. Our numerical simulations, fig.~\ref{snr-spike-psd}(b),
fully confirms this prediction. Conventional stochastic resonance
therefore occurs only for large membrane patches beyond optimal size, and reaches
saturation
in the limit $S\to\infty$ (limit of the deterministic Hodgkin-Huxley model). 
Thus, the observed
biological SR \cite{Douglass93,Levin96} is rooted in the 
collective properties of
large ion channels arrays, where ion channels are  
globally coupled via the common membrane
potential $V(t)$.

In conclusion,  we have investigated stochastic and coherence resonance in
a noisy generalization of the Hodgkin-Huxley model for
excitable biological cell membrane patches.
The spontaneous fluctuations of the membrane conductivity due to
the individual channel dynamics
has  systematically
been taken into account. 
 We have shown that the excitable membrane patches
with an observation area around $S\approx 1{\rm \mu m^{2}}$ exhibit a
rhythmic spiking activity optimized by everpresent internal noise.
In other words, the collective dynamics of globally coupled
ion channels become more ordered solely due to {\it internal} noise.
This new effect can be regarded as the 
{\it intrinsic coherence resonance phenomenon}; it  presents a 
first important result of our work. This very finding has also been confirmed
independently within a different approach by Jung and Shuai \cite{Jung} 
(see the accompanying paper). A second main result of this study refers to the 
phenomenon of {\it intrinsic SR}; thereby the internal noise {\it alone} gives rise to
a SR behavior, see fig.~\ref{snr-spike-psd}(a).   Conventional SR versus the external noise intensity
takes place as well for sufficiently large membrane patches where the internal noise strength
alone is not yet at its  optimal value. We hence conclude that
 observed biological SR likely is rooted in the {\it collective} properties
of globally coupled ion channel assemblies. 

The authors gratefully 
acknowledge the support of this work by the Deutsche
Forschungsgemeischaft, SFB 486 {\it Manipulation
  of matter on the nanoscale}, Project A10.
Moreover, we acknowledge most helpful and constructive discussions with Peter Jung.

\end{document}